\documentclass[english,aps,pra,twocolumn,superscriptaddress,showpacs]{revtex4}
\usepackage{amsmath}
\usepackage[T1]{fontenc}
\usepackage[latin1]{inputenc}
\usepackage{amsmath,amsfonts}
\usepackage{epsf}
\usepackage{graphicx}
\usepackage{amssymb}
\usepackage{babel}
\usepackage{color}
\usepackage{ifpdf}
\usepackage{comment}
\usepackage[normalem]{ulem}

\def\be{\begin{equation}}
\def\ee{\end{equation}}
\def\bea{\begin{eqnarray}}
\def\eea{\end{eqnarray}}

\def\nn{\nonumber}

\makeatletter

\begin{document}

\title{Exotic pairing states in a Fermi gas with three-dimensional spin-orbit coupling}

\author{Xiang-Fa Zhou}
\affiliation{Key Laboratory of Quantum Information, University of Science and Technology of China, CAS, Hefei, Anhui, 230026, People's Republic of China}
\author{Guang-Can Guo}
\affiliation{Key Laboratory of Quantum Information, University of Science and Technology of China, CAS, Hefei, Anhui, 230026, People's Republic of China}
\author{Wei Zhang}
\email{wzhangl@ruc.edu.cn}
\affiliation{Department of Physics, Renmin University of China, Beijing 100872, People's Republic of China}
\author{Wei Yi}
\email{wyiz@ustc.edu.cn}
\affiliation{Key Laboratory of Quantum Information, University of Science and Technology of China, CAS, Hefei, Anhui, 230026, People's Republic of China}

\begin{abstract}
We investigate properties of exotic pairing states in a three-dimensional Fermi gas with three-dimensional spin-orbit coupling and an effective Zeeman field. The interplay of spin-orbit coupling, effective Zeeman field and pairing can lead to first-order phase transitions between different phases, and to interesting nodal superfluid states with gapless surfaces in the momentum space. We then demonstrate that pairing states with zero center-of-mass momentum are unstable against Fulde-Ferrell-Larkin-Ovchinnikov (FFLO) states, with a finite center-of-mass momentum opposite to the direction of the effective Zeeman field. Unlike conventional FFLO states, these FFLO states are induced by the coexistence of spin-orbit coupling and Fermi surface deformation, and have intriguing features like first-order transitions between different FFLO states, nodal FFLO states with gapless surfaces in momentum space, and exotic fully gapped FFLO states. With the recent theoretical proposals for realizing three-dimensional spin-orbit coupling in ultracold atomic gases, our work is helpful for the future experimental studies, and provides valuable information for the general understanding of pairing physics in spin-orbit coupled fermionic systems.
\end{abstract}
\maketitle
\section{Introduction}
Ultracold atomic gases have been considered as ideal platforms for the quantum simulation of interesting models in a variety of different physical contexts, ranging from condensed matter physics, to nuclear physics \cite{blochreview,ketterlereview,chengchinreview}. With powerful tools like Feshbach resonance, optical lattice etc., strongly correlated models can be studied in this novel type of experimental systems. Of particular importance is the possibility to simulate models that have no counterparts in other systems. In the past decade or so, interesting phenomena like BCS-BEC crossover, superfluid to Mott insulator transition in a Bose-Hubbard model etc., which do not occur naturally in condensed matter systems, have been investigated experimentally using ultracold atom gases \cite{blochreview,ketterlereview}.

Recently, by engineering the atom-laser coupling, synthetic spin-orbit coupling (SOC) has been realized experimentally in ultracold atoms \cite{gauge2exp,fermisocexp1,fermisocexp2,shuaiexp}. As SOC is considered to play a key role in systems with interesting properties like quantum spin Hall effect or topological phases \cite{kanereview}, the experimental realization of SOC in ultracold atoms has greatly extended the possibility of quantum simulation in these systems. A considerable amount of theoretical and experimental efforts have since been dedicated to the characterization of rich physics in spin-orbit coupled atomic gases \cite{zhang,sato,hzsocbec,cplwu,caizipra,soc3,chuanwei,soc4,soc6,iskin,thermo,2d2,2d1,melo,wmliu,helianyi,wy2d,xiaosen,wypolaron,chuanweifflo,carlosnistsoc,puhantwobody,shenoy,iskinnistsoc,wyfflo,melosupp1,melosupp2,hufflo,puhan3dsoc}. Most of these studies have focused on the Rashba or Dresselhaus SOC, which also exist in solid state systems; or on the NIST SOC, which can be readily implemented with present technology \cite{gauge2exp,fermisocexp1,fermisocexp2,shuaiexp,xiongjun}. More exotic forms of SOC, such as a three-dimensional (3D) SOC, which have not been observed in natural condensed matter systems, have also
attracted some attention recently in bosonic systems \cite{3dsocbec1,3dsocbec2,3dsocbec3,3dsocbec4,3dsocbec5,3dsoc2body1,3dsoc2body2}. With the recent proposals of implementing 3D SOC in ultracold atomic gases \cite{3dsoc1,3dsoc2,3dsoc3,3dsoc4,3dsoc5}, it is hopeful that interesting physics like 3D topological insulators \cite{kanereview}, Weyl semi-metals \cite{weyl}, or pairing physics in a Fermi gas etc. \cite{puhan3dsoc}, could be investigated in these systems.

Indeed, a particularly interesting problem that has been under intensive theoretical study lately is the pairing superfluidity in an attractively interacting Fermi gas with SOC and effective Zeeman fields. In a two-dimensional (2D) Fermi gas with Rashba SOC and axial Zeeman field, it has been shown that a topological superfluid state can be stabilized, which supports topologically protected edge states and Majorana zero modes at the core of vortex excitations \cite{zhang,sato,soc1,wy2d}. While for a 2D Fermi gas with the experimentally available NIST SOC or for a 3D Fermi gas with Rashba SOC, nodal superfluid states with gapless excitations exist \cite{chuanwei,iskin,thermo,carlosnistsoc,iskinnistsoc,wyfflo,melosupp1,melosupp2}. Furthermore, it has been suggested that for a 2D Fermi gas under NIST SOC and cross Zeeman fields, finite center-of-mass momentum Fulde-Ferrell-Larkin-Ovchinnikov (FFLO) states become the ground state of the system \cite{wyfflo}. These FFLO states are different from the conventional FFLO pairing states in a polarized Fermi gas in that they are driven by SOC-induced spin mixing and transverse-field-induced Fermi surface deformation \cite{chuanweifflo,wyfflo,hufflo,puhan3dsoc}. A natural question that follows is whether such an exotic pairing state is a unique feature of systems with SOC and Fermi surface asymmetry.

In this work, we study the pairing states in a 3D attractively interacting Fermi gas with 3D SOC and an effective axial Zeeman field \cite{footnote1}. As previous studies with 3D SOC have mostly focused on bosonic systems, our work provides a starting point for the understanding of the effects of 3D SOC in a fermionic system. Additionally, the Fermi surfaces of such a system are asymmetric along the direction of the effective Zeeman field. Hence it is an ideal system for the study of pairing states under SOC and Fermi surface asymmetry. To get a qualitative understanding of the pairing physics, we focus on the zero-temperature phase diagram on the mean-field level.

For the zero center-of-mass momentum pairing states, we find that exotic nodal superfluid states exist with either two or four closed gapless surfaces in momentum space. As the anisotropy of the 3D SOC along the axial direction increases, these gapless surfaces shrink gradually, and will eventually collapse into gapless points as the 3D SOC is reduced to a Rashba form. Therefore, these nodal superfluid states can be seen as the counterparts of the nodal superfluid states in a 3D Fermi gas with Rashba SOC and an effective Zeeman field \cite{thermo}. We also find that the interplay between SOC, effective Zeeman field and pairing in the system can lead to first-order phase boundaries between different superfluid states, which implies the existence of a phase separated state in a uniform gas and spatial phase separation in a trapped gas. These findings provide a valuable starting point for understanding finite center-of-mass momentum pairing states.

When the finite center-of-mass momentum pairing is taken into account, the phase diagram is qualitatively modified. Similar to the effects of a transverse field in the NIST scheme, the effective axial Zeeman field in combination with the 3D SOC makes the Fermi surface asymmetric. As a result, pairing states with zero center-of-mass momentum become unstable against FFLO states, with the center-of-mass momentum of the pairs determined by the Fermi surface asymmetry \cite{wyfflo}. These FFLO states are characteristic of pairing states in the presence of spin-orbit coupling and Fermi surface deformation, which induce a shift of the local minima in the thermodynamic potential landscape onto the plane of finite center-of-mass momentum. Consequently, these FFLO states retain many features of their zero center-of-mass momentum counterparts, e.g. the existence of first-order boundaries between different FFLO states, the stabilization of nodal FFLO states with gapless surfaces in momentum space, and the presence of a continuous transition between the gapless and a fully gapped FFLO state. As similar scenarios have also been reported in Fermi gases with NIST SOC in different dimensions \cite{wyfflo,hufflo}, we expect that these should be the unique features for pairing states under SOC and Fermi surface asymmetry, and should be independent of the concrete form of SOC.

We stress that the FFLO state considered in this work is in fact the Fulde-Ferrell (FF) state, i.e., pairing state with a single center-of-mass momentum $\mathbf{Q}$. More generally, one should also consider the Larkin-Ovchinnikov (LO) state, where the center-of-mass momentum of the pairing state has both $\mathbf{Q}$ and $-\mathbf{Q}$ components. The stability of the LO state in a spin-orbit coupled system is a subtle issue and can be investigated for example, by solving the Bogoliubov-de Gennes equation \cite{bdg1,bdg2}.

The paper is organized as follows: in Sec. I, we present our mean-field formalism; in Sec II, we first discuss the phase diagram when only zero center-of-mass momentum pairing states are considered; we take finite center-of-mass pairing states into account in Sec. III, where we also discuss the general physical picture of pairing under SOC and Fermi surface asymmetry. Finally, we summarize in Sec. IV.

\section{Model}
We consider a 3D uniform two component Fermi gas with an axially symmetric 3D spin-orbit coupling $\mathbf{\sigma}_{\perp}\cdot
\mathbf{k}_{\perp}+\gamma \sigma_z k_z$, where $\sigma_i$ ($i=x,y,z$) are the Pauli matrices, and the parameter $\gamma$ depicts the anisotropy of the 3D SOC. The system is further subject to an effective Zeeman field along the $\hat{z}$ direction. The Hamiltonian of the system can be written as
\begin{eqnarray}
H &=& \sum_{\mathbf{k}}(\epsilon_{\mathbf{k}}-\mu)(a^{\dag}_{\mathbf{k}} a_{\mathbf{k}} +
b^{\dag}_{\mathbf{k}}b_{\mathbf{k}}) -\frac{h}{2} \sum_{\mathbf{k}}(a^{\dag}_{\mathbf{k}} a_{\mathbf{k}}
-b^{\dag}_{\mathbf{k}}b_{\mathbf{k}})  \nn \\
 &+&\sum_{\mathbf{k}}\alpha  k \sin \theta_{\mathbf{k}}(e^{-i \phi_{\mathbf{k}}} a^{\dag}_{\mathbf{k}} b_{\mathbf{k}}
+ e^{i \phi_{\mathbf{k}}}b^{\dag}_{\mathbf{k}}a_{\mathbf{k}})\nonumber\\
&+& \frac{U}{V} \sum_{\mathbf{k},\mathbf{k}',\mathbf{q}} a^{\dag}_{\mathbf{k}+\frac{\mathbf{q}}{2}}
b^{\dag}_{-\mathbf{k}+\frac{\mathbf{q}}{2}} b_{-\mathbf{k}'+\frac{\mathbf{q}}{2}} a_{\mathbf{k}'+\frac{\mathbf{q}}{2}}\nonumber\\
&+& \gamma \sum_{\mathbf{k}}\alpha k\cos \theta_{\mathbf{k}} (a^{\dag}_{\mathbf{k}} a_{\mathbf{k}}
-b^{\dag}_{\mathbf{k}}b_{\mathbf{k}}),\label{Heqn}
\end{eqnarray}
where $\mathbf{k}=k(\sin \theta_{\mathbf{k}} \cos \phi_{\mathbf{k}}, \sin \theta_{\mathbf{k}} \sin \phi_{\mathbf{k}}, \cos \theta_{\mathbf{k}})$, $V$ is the quantization volume in 3D , $\epsilon_{\mathbf{k}}=\hbar^2k^2/(2m)$, $a_{\mathbf{k}}$ ($a^{\dag}_{\mathbf{k}}$) and $b_{\mathbf{k}}$ ($b^{\dag}_{\mathbf{k}}$) represent the annihilation (creation) operators for atoms of different spin species, $\mu$
is the chemical potential, $h$ is the effective Zeeman field, $\alpha$
is the spin-orbit coupling strength. We have assumed $s$-wave contact interaction between atoms of different spin species, where the bare interaction rate $U$ should be renormalized following the standard relation \cite{zhangpeng}
\begin{equation}
\frac{1}{U}=\frac{1}{U_p}-\frac{1}{V}\sum_{\mathbf{k}}\frac{1}{2\epsilon_{\mathbf{k}}}.
\end{equation}
The physical interaction rate $U_p$ is defined as $U_p=4\pi\hbar^2 a_s/m$, with $a_s$ the s-wave scattering length between the two spin species, which can be tuned via the Feshbach resonance technique. We note that while the SOC becomes isotropic when the anisotropy parameter $\gamma=1$, it reduces to the Rashba type SOC for $\gamma=0$.

The Hamiltonian (\ref{Heqn}) can be diagonalized under the mean-field approximation, for which we define the FF pairing order parameter \cite{fflo}: $\Delta_{\textbf{Q}} =U/V \sum_{\mathbf{k}} \langle b_{-\mathbf{k}+\textbf{Q}/2} a_{\mathbf{k}+\textbf{Q}/2} \rangle
$. The resulting effective Hamiltonian can be arranged into a matrix form under the basis $\psi_{\mathbf{k}+\textbf{Q}/2}=(a_{\mathbf{k}+\textbf{Q}/2}, b_{\mathbf{k}+\textbf{Q}/2}, a^{\dag}_{-\mathbf{k}+\textbf{Q}/2}, b^{\dag}_{-\mathbf{k}+\textbf{Q}/2})^T$
\begin{eqnarray}
H_{\text{eff}}&=&\frac{1}{2}\sum_{\mathbf{k}} \psi^{\dag}_{\mathbf{k}+\textbf{Q}/2} M^{\textbf{Q}}_{\mathbf{k}} \psi_{\mathbf{k}+\textbf{Q}/2}
+ \sum_{\mathbf{k}}(\epsilon_{\mathbf{k}+\textbf{Q}/2}-\mu)\nonumber\\
&-& \frac{V}{U}
|\Delta_{\textbf{Q}}|^2, \label{Hmatrix}
\end{eqnarray}
 with
\begin{eqnarray}
M^{\textbf{Q}}_{\mathbf{k}}=\left (
\begin{array}{cc}
 H_0(\mathbf{k}+\textbf{Q}/2) &  i \Delta_{\textbf{Q}} \sigma_y \\
-i \Delta_{\textbf{Q}} \sigma_y & - H^*_0(-\mathbf{k}+\textbf{Q}/2)
\end{array}
\right ).
\end{eqnarray}
Here, $H_0(\mathbf{k}) = (\epsilon_{\mathbf{k}}-\mu)I -(\frac{h}{2}-\alpha \gamma k_z) \sigma_z +
\alpha \mathbf{\sigma}_{\perp} \cdot \mathbf{k}_{\perp}$, where $I$ is the identity matrix. Diagonalizing the effective Hamiltonian (\ref{Hmatrix}), we may then evaluate the zero-temperature thermodynamic potential
\begin{eqnarray}
\Omega&=&\frac{1}{4}
\sum_{\mathbf{k},\nu,\lambda}(|E_{\mathbf{k},\nu}^{\lambda}|-E_{\mathbf{k},\nu}^{\lambda})+\sum_{\mathbf{k}}(\epsilon_{\mathbf{k}+\textbf{Q}/2}-\mu+\frac{|\Delta_{\textbf{Q}}|^2}{2\epsilon_{\mathbf{k}}})\nonumber\\
&-&\frac{V}{U_p}|\Delta_{\textbf{Q}}|^2.\label{thermoeqn}
\end{eqnarray}
Here, the quasi-particle (hole) dispersion $E_{\mathbf{k},\nu}^{\lambda}$ ($\nu,\lambda=\pm$) are the eigenvalues of the matrix $M^{\mathbf{Q}}_{\mathbf{k}}$ in Eq. (\ref{Hmatrix}). Without loss of generality, we assume $h>0$, $\Delta_0\equiv\Delta$, and $\Delta_Q$ to be real throughout the work.

Typically, the ground state of the system can be found by solving the gap equation $\partial \Omega/\partial \Delta_{\mathbf{Q}}=0$ and the number equation $n=-(1/V)\partial\Omega/\partial \mu$ simultaneously, where $n$ is the total particle number density. However, the competition between pairing and the effective Zeeman field leads to a double-well structure in the thermodynamic potential for both the $Q=0$ and the finite $\mathbf{Q}$ cases. Therefore, to get the correct ground state of a uniform gas, one must explicitly take into account of the possibility of phase separation between the phases that correspond to the degenerate local minima of the thermodynamic potential \cite{preview}. Indeed, for calculations in a canonical ensemble where the total particle number is fixed, a phase-separated state must be taken into consideration when minimizing the Helmholtz free energy. As we will show in Sec. \ref{sec5}, this is reflected in the fact that for certain total number densities, the number equation and the gap equation cannot be solved simultaneously, unless a phase separated state is considered. On the other hand, for a fixed chemical potential, one does not need to solve the number equation, hence there is no phase separation. In this work, we fix the chemical potential $\mu$ and look for the global minimum of the thermodynamic potential (\ref{thermoeqn}) as a function of the pairing order parameter $\Delta_{\mathbf{Q}}$ and center-of-mass momentum $\mathbf{Q}$. Considering the fact that different spin states are mixed in the presence of SOC, the phase diagram obtained from this algorithm can be easily connected to that of a uniform system with a fixed total particle number. Besides, the global minimum of the thermodynamic potential for a fixed chemical potential also corresponds to the local ground state in a trapped gas under the local density approximation (LDA). The total particle number in the trap can then be evaluated by integrating the local number densities from the trap center to its edge, i.e. by integrating the number equation with decreasing chemical potentials. Thus, it is straightforward to reveal the phase structure in slow-varying potential traps from the resulting phase diagram under LDA.

\section{Pairing states with zero center-of-mass momentum}

\begin{figure}
    \includegraphics[width=4.2cm]{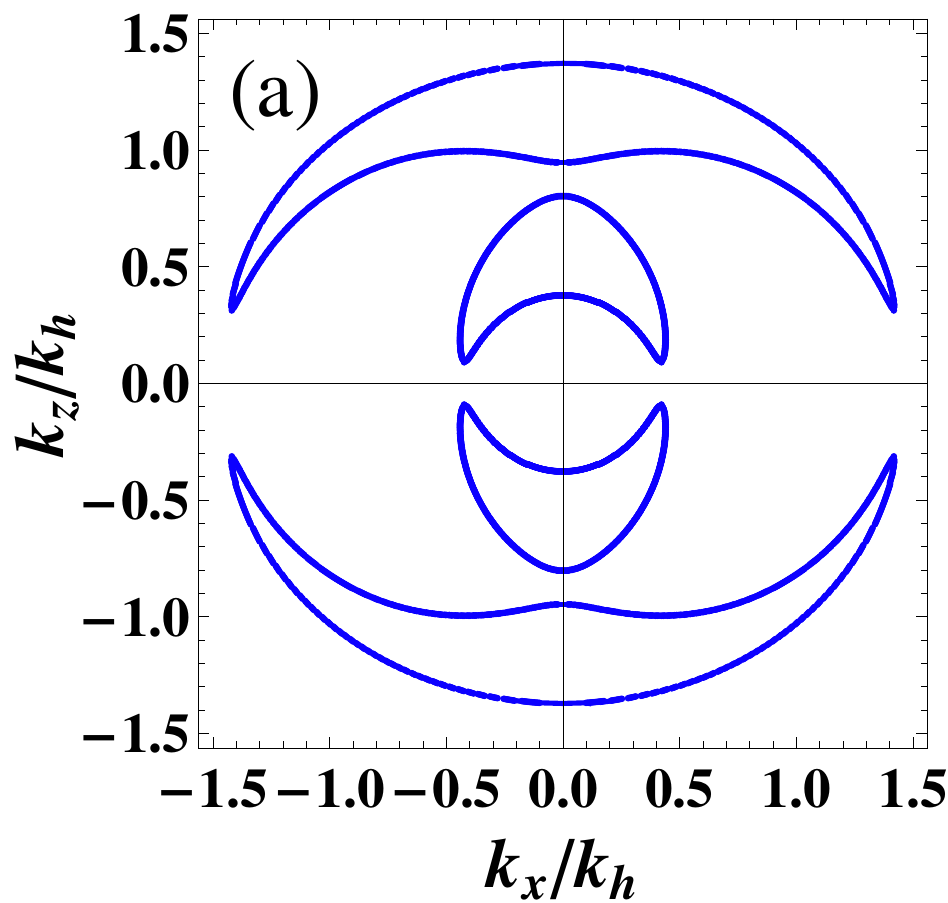}
    \includegraphics[width=4.2cm]{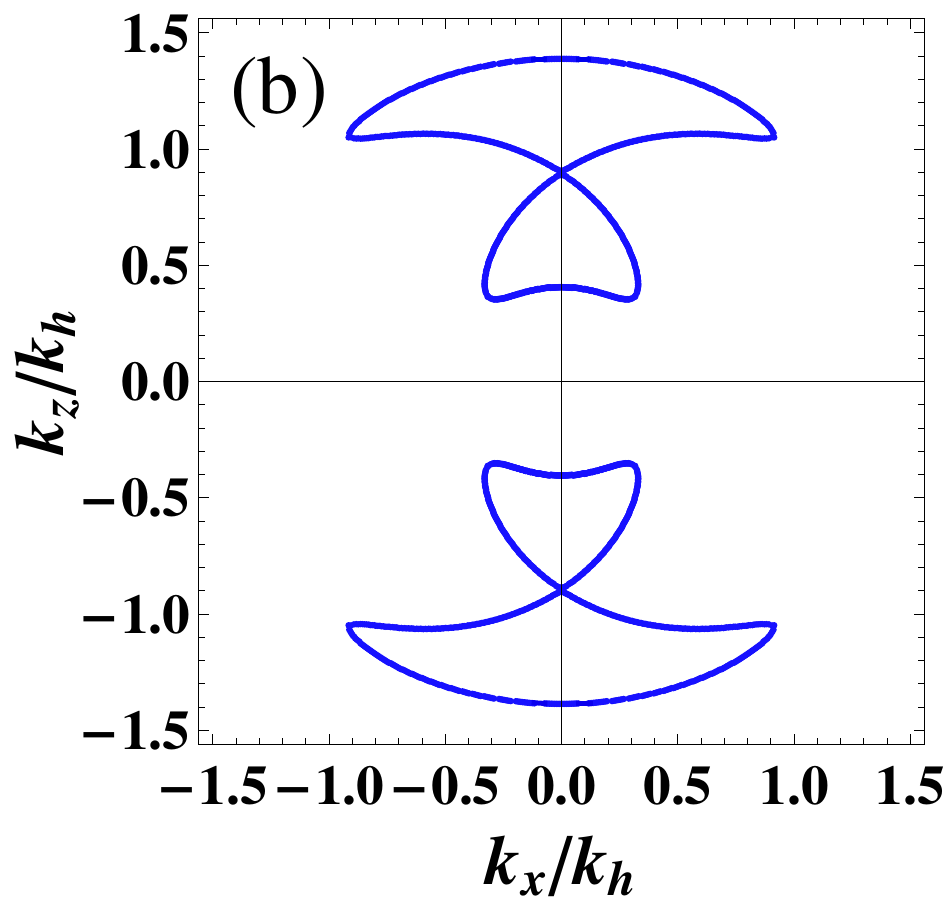}
    \includegraphics[width=4.2cm]{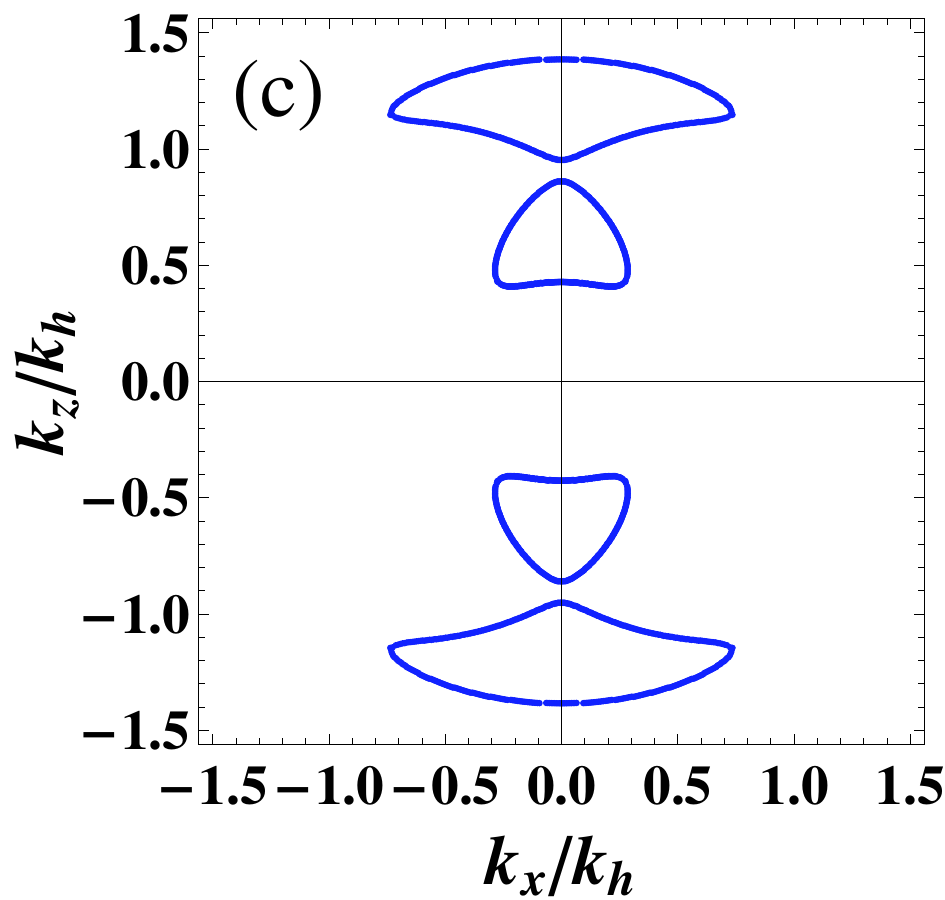}
    \includegraphics[width=4.2cm]{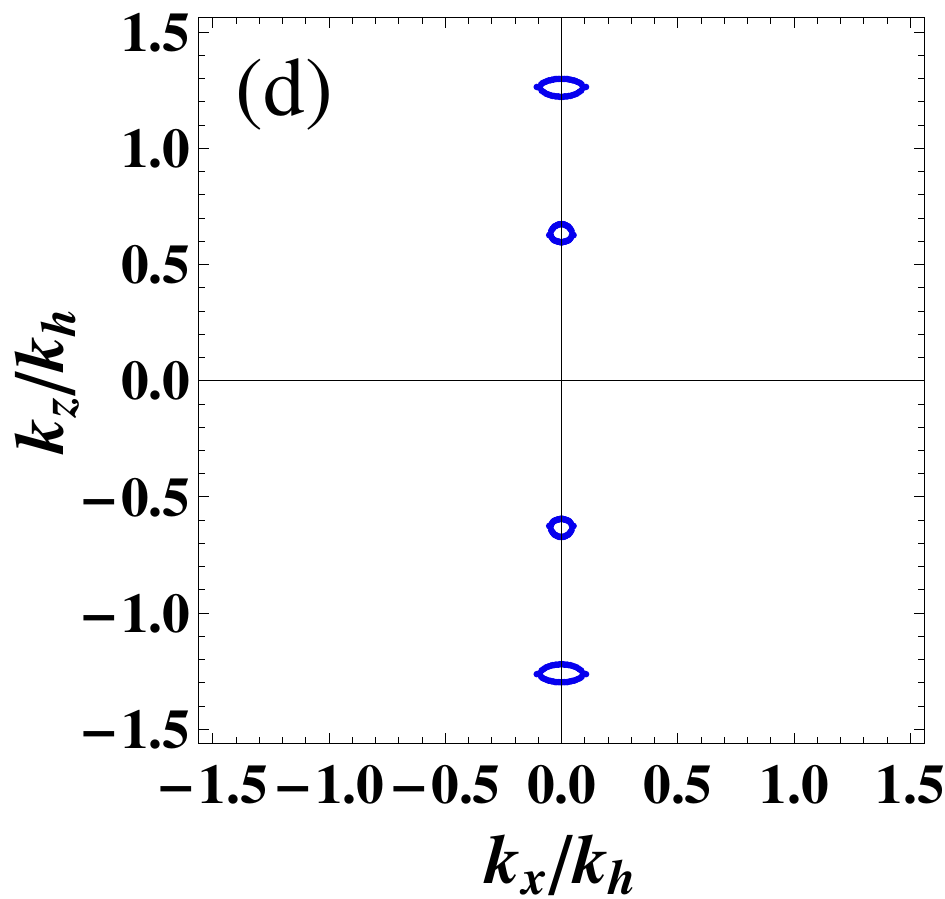}
    \includegraphics[width=4.2cm]{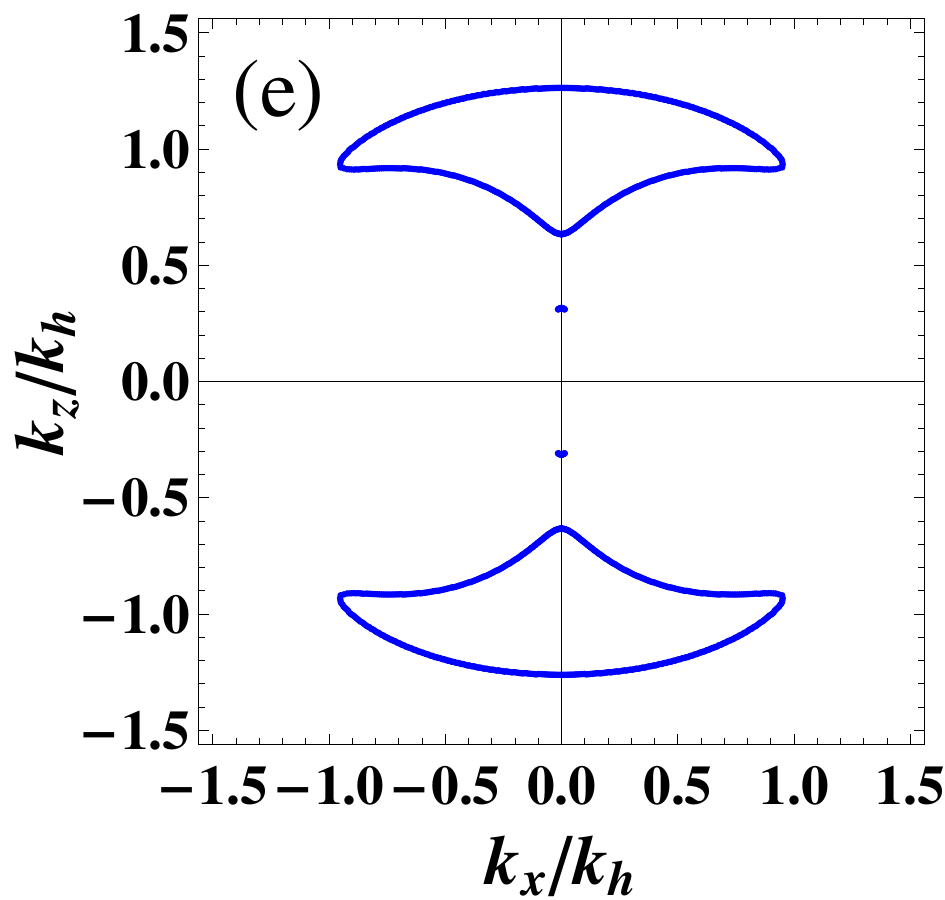}
    \includegraphics[width=4.2cm]{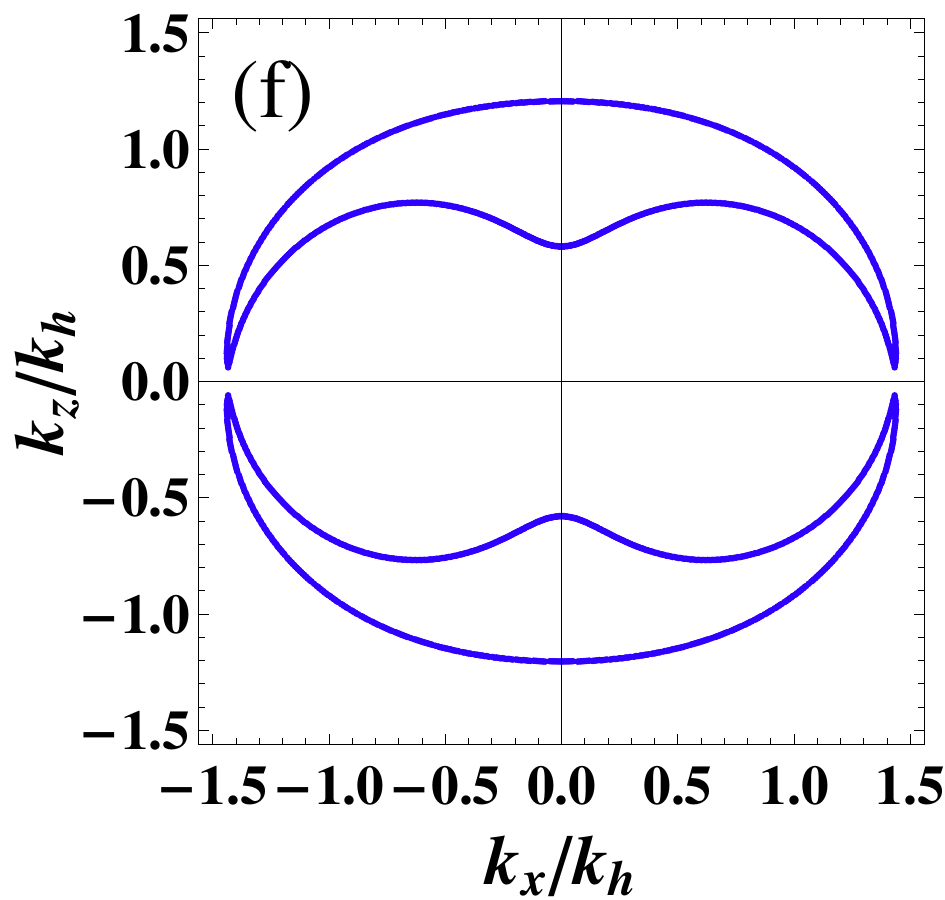}
\caption{(Color online) Gapless contours of the ground states with $Q=0$ in the $k_x$--$k_z$ plane ($k_y=0$) for $\gamma=0.5$ and $(k_ha_s)^{-1}=-1$, with the parameters : (a) $\alpha k_h/h=0.85$, $\mu/h=0.8$, $\Delta/h\sim0.06$; (b) $\alpha k_h/h=1$, $\mu/h=0.8$, $\Delta/h\sim0.25$; (c) $\alpha k_h/h=1.05$, $\mu/h=0.8$, $\Delta/h\sim0.31$; (d) $\alpha k_h/h=1.25$, $\mu/h=0.8$, $\Delta/h\sim0.49$; (e) $\alpha k_h/h=1.26$, $\mu/h=0.35$, $\Delta/h\sim0.22$; (f) $\alpha k_h/h=1.25$, $\mu/h=0.2$, $\Delta/h\sim0.01$. Note the gapless contours have axial symmetry around $z$-axis.  The effective Zeeman field $h$ is taken to be the unit of energy, while the unit of momentum $k_h$ is defined through $\hbar^2k_h^2/2m=h$.
}\label{fig1contour}
\end{figure}

\begin{figure}
    \includegraphics[width=8cm]{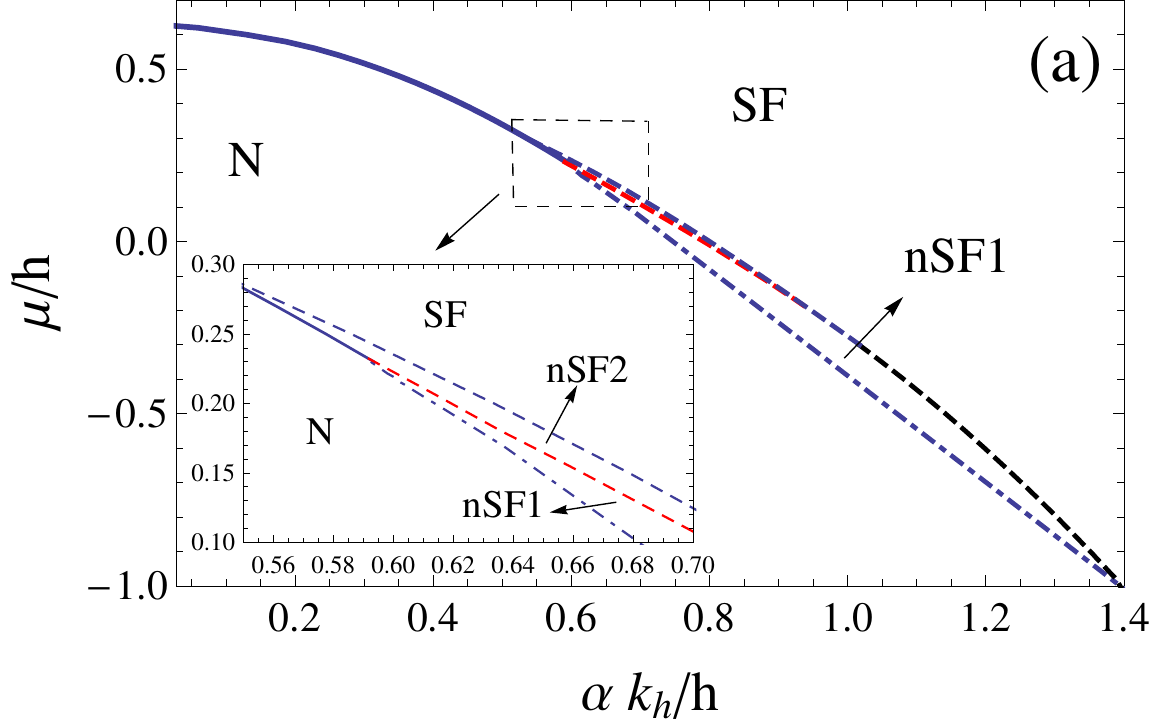}
    \includegraphics[width=8cm]{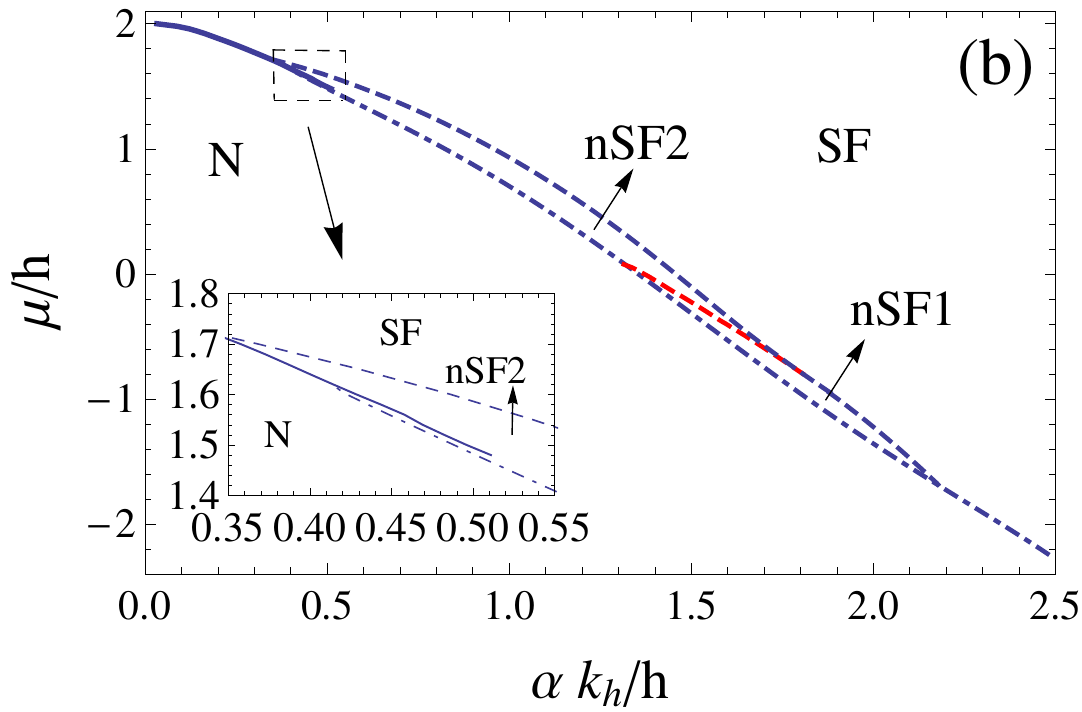}
\caption{(Color online) Phase diagram obtained in the $\mu$--$\alpha$ plane with (a) $(k_ha_s)^{-1}=0$, and (b) $(k_ha_s)^{-1}=-1$, respectively. The bold solid curves represent the first-order boundaries, the thin dashed curves represent continuous phase boundaries, and dash-dotted curves represent the normal state threshold $\Delta/h=10^{-3}$. For both figures, we have set $\gamma=1$. }\label{fig2phase}
\end{figure}

We start by analyzing the pairing states with zero center-of-mass momentum, which will provide us with valuable information for the FFLO pairing states that we will discuss in the following section. For $Q=0$, an analytical expression for the quasi-particle (hole) dispersions is typically not available, except for those along the $k_z$ axis, which can be written as
\begin{equation}
E^{\lambda}_{k_{\perp}=0,\pm}=-\lambda\big[ \frac{h}{2} \pm \sqrt{(\epsilon_{\mathbf{k}}-\mu+\lambda \alpha \gamma k_z)^2+\Delta^2}
\big].\label{dispersionQ0}
\end{equation}
Apparently, two of the branches $E^{\lambda}_{\mathbf{k},-}$ can cross zero, leading to nodal superfluid states with gapless excitations. As the dispersion spectra have the symmetry $E^{+}_{k_z,\nu}=-E^{-}_{-k_z,\nu'}$, the gapless points on the $k_z$ axis must be symmetric with respect to the origin. Furthermore, we find that the gapless points in momentum space typically form closed surfaces that are axially symmetric with respect to the $k_z$ axis. Typical gapless contours in the $k_x$--$k_z$ plane are shown in Fig. (\ref{fig1contour}), where we demonstrate the evolution of the gapless surfaces as the parameters are tuned. It is clear that there can be two or four disconnected gapless surfaces in momentum space, for which we may define different nodal superfluid states. The number of these closed surfaces can be deduced by counting gapless points on the $k_z$ axis. From Eq. (\ref{dispersionQ0}), we know that the possible gapless points are
$k_z^0=\lambda \Big [ \alpha \gamma m /\hbar^2 +\nu\sqrt{\alpha^2\gamma^2m^2/(2\hbar^4) + 2m\mu/\hbar^2 +\chi 2m/\hbar^2\sqrt{h^2/4-\Delta^2}}
\Big]$. Typically, there are eight gapless points when $\mu +\alpha^2\gamma^2 m/(2\hbar^2)>\sqrt{h^2/4-\Delta^2}$ with $\lambda,\nu,\chi=\pm$; four gapless points
when $\left|\mu +\alpha^2\gamma^2 m/(2\hbar^2)\right|<\sqrt{h^2/4-\Delta^2}$ with $\lambda,\nu\pm,\chi=+$; and no gapless points otherwise.
The three different cases correspond to nodal superfluid state with four closed gapless surfaces (nSF2), with two closed gapless surfaces (nSF1), and the fully gapped superfluid state (SF), respectively. In the case of $\gamma=0$, these gapless surfaces collapse into gapless points on the $k_z$-axis, recovering the results of a 3D Fermi gas with Rashba SOC \cite{thermo}.

\begin{figure}
\includegraphics[width=8cm]{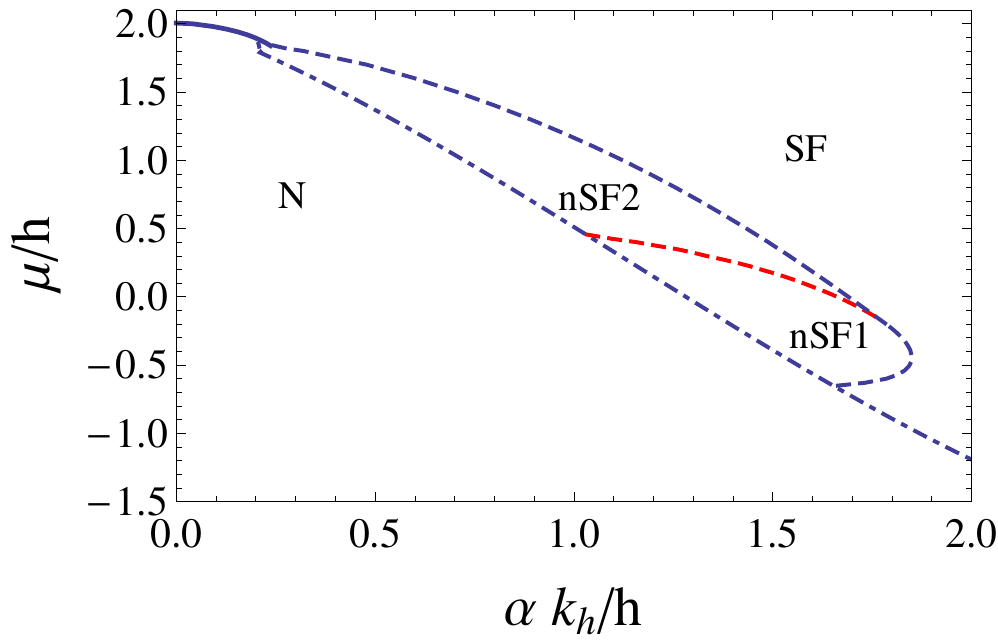}
\caption{(Color online) Phase diagram in the $\mu$--$\alpha$ plane with $\gamma=0.5$, $(k_h a_s)^{-1}=-1$. The bold solid curves represent the first-order boundaries, the thin dashed curves represent continuous phase boundaries, and dash-dotted curves represent the normal state threshold $\Delta/h=10^{-3}$.}\label{fig3phase}
\end{figure}

With the boundaries between different superfluid states fixed, we may then map out the typical phase diagrams on the $\mu$--$\alpha$ plane. As discussed in the previous section, the phase diagram reflects the phase structure of a trapped gas under LDA. The phase diagram for a homogeneous system with fixed particle number density can be easily extracted from the resulting $\mu$--$\alpha$ phase diagram by evaluating the particle number at the phase transition lines (c.f. Sec. IV). In Fig. \ref{fig2phase}, we show the phase diagrams for pairing states with $Q=0$ and under an isotropic SOC ($\gamma=1$). As we have anticipated, there exist two different nodal superfluid states, with either two (nSF1) or four (nSF2) gapless surfaces in momentum space. The stability region of the nodal superfluid states increase slightly toward the BCS side. Another prominent feature is the existence of first-order phase boundaries on the phase diagram. Similar first-order boundaries have been reported for Fermi gases in various dimensions and with different forms of SOC and effective Zeeman fields \cite{thermo,wy2d,xiaosen,wyfflo}. Apparently, these first-order boundaries are due to the competition between the double wells in the thermodynamic potential, and are rooted in the complex interplay between SOC, Zeeman field and pairing. As a consequence, one needs to consider explicitly the possibility of a phase-separated state in a uniform gas. We note that due to numerical uncertainties, the continuous boundary of the normal state is determined by the threshold $\Delta/h=10^{-3}$.

To demonstrate the influence of anisotropic SOC on the pairing states, we show in Fig. \ref{fig3phase} a typical phase diagram for $\gamma=0.5$. It appears that the SOC anisotropy enhances the stability of the nodal superfluid states, and destabilizes the phase-separated state as it makes the first-order boundary shorter. In the case of $\gamma=0$, we recover the corresponding phase diagram of a 3D Fermi gas with Rashba SOC and an axial Zeeman field, where large stability regions for the nodal superfluid states can be identified. The first-order boundary though would persist in the $\gamma=0$ limit \cite{thermo}.

\section{FFLO pairing states under 3D SOC and Zeeman field}\label{sec5}

\begin{figure}
\includegraphics[width=8cm]{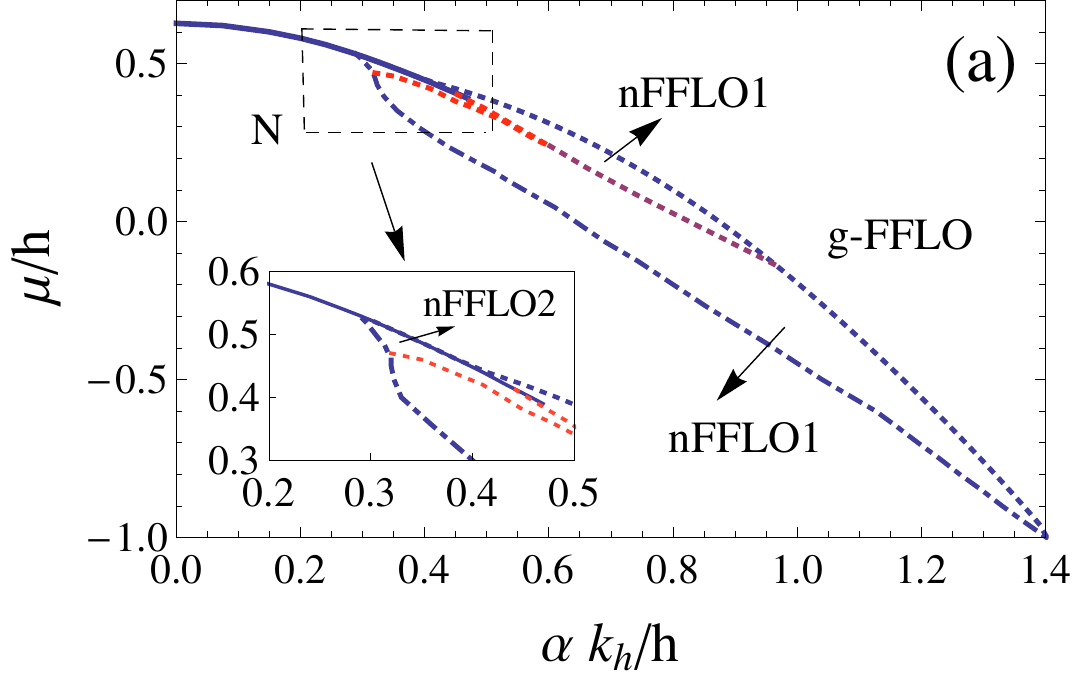}
\includegraphics[width=8cm]{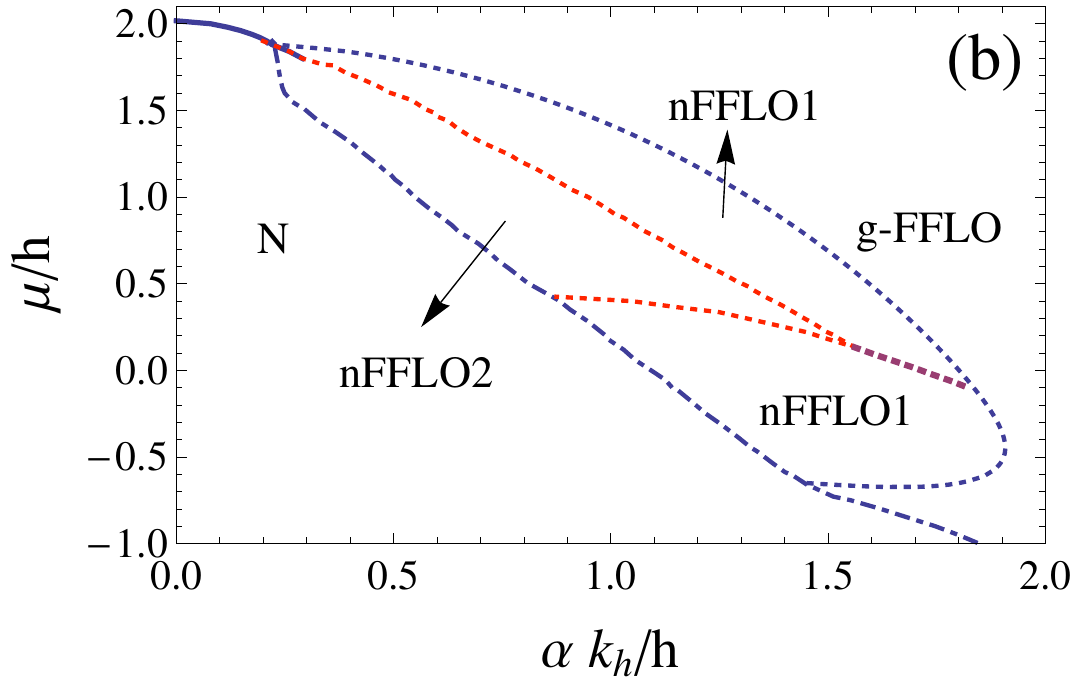}
\caption{(Color online) Phase diagram for FFLO states in the $\mu$--$\alpha$ plane, with (a) $(k_h a_s)^{-1}=0$, $\gamma=1$ and (b)  $(k_h a_s)^{-1}=-1$, $\gamma=0.5$. The bold solid curves represent the first-order boundaries, the dashed curves represent the continuous boundaries between the nFFLO1 and the nFFLO2, and the g-FFLO state, and the dash-dotted curve represents a normal state threshold with $\Delta_{\mathbf{Q}}/h=10^{-3}$.}\label{figfflophase}
\end{figure}

\begin{figure}
\includegraphics[width=4.2cm]{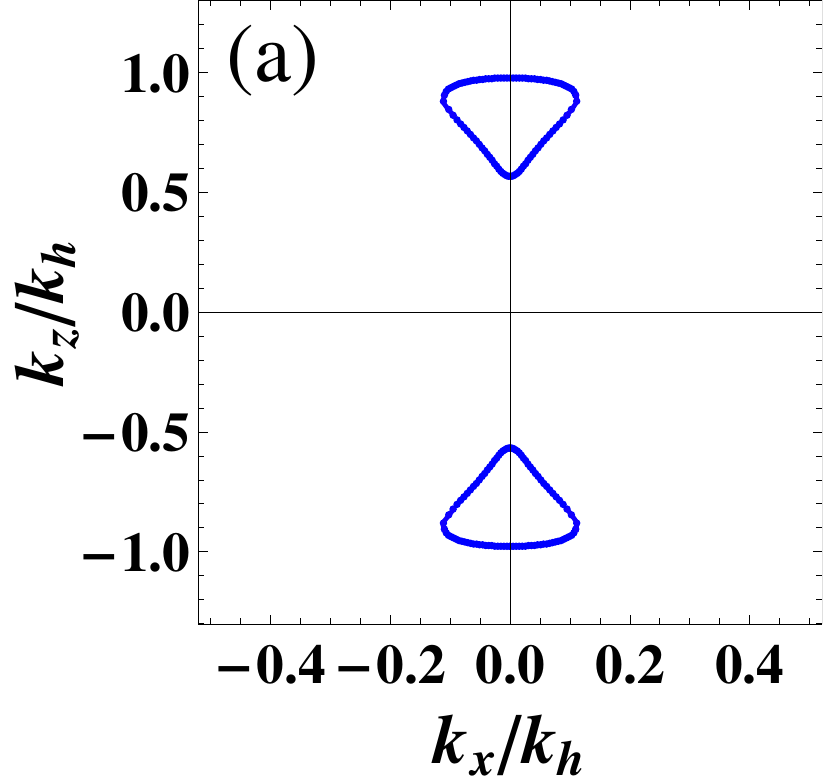}
\includegraphics[width=4.2cm]{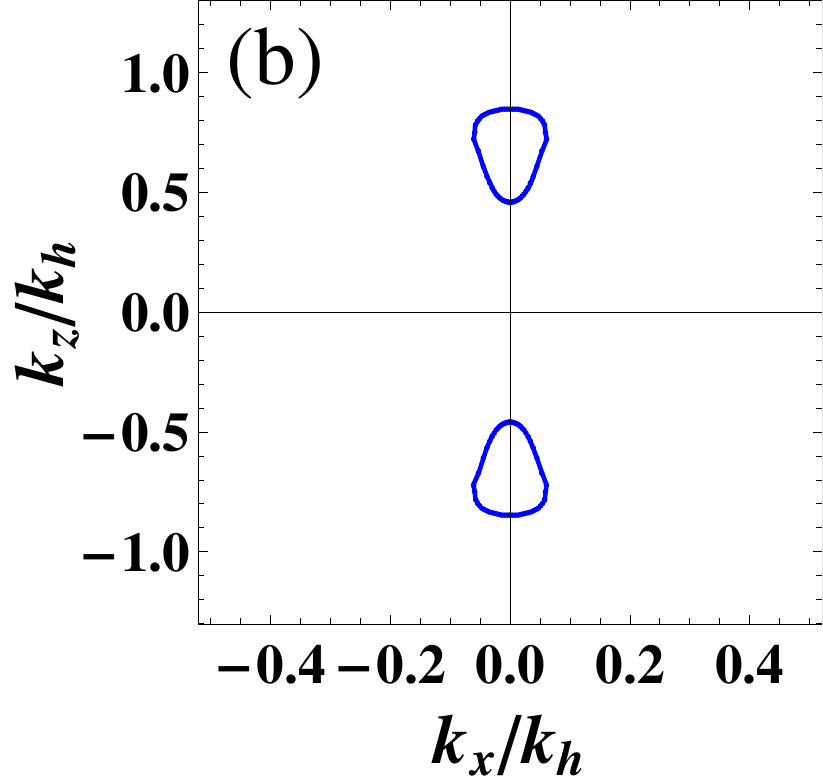}
\includegraphics[width=4.2cm]{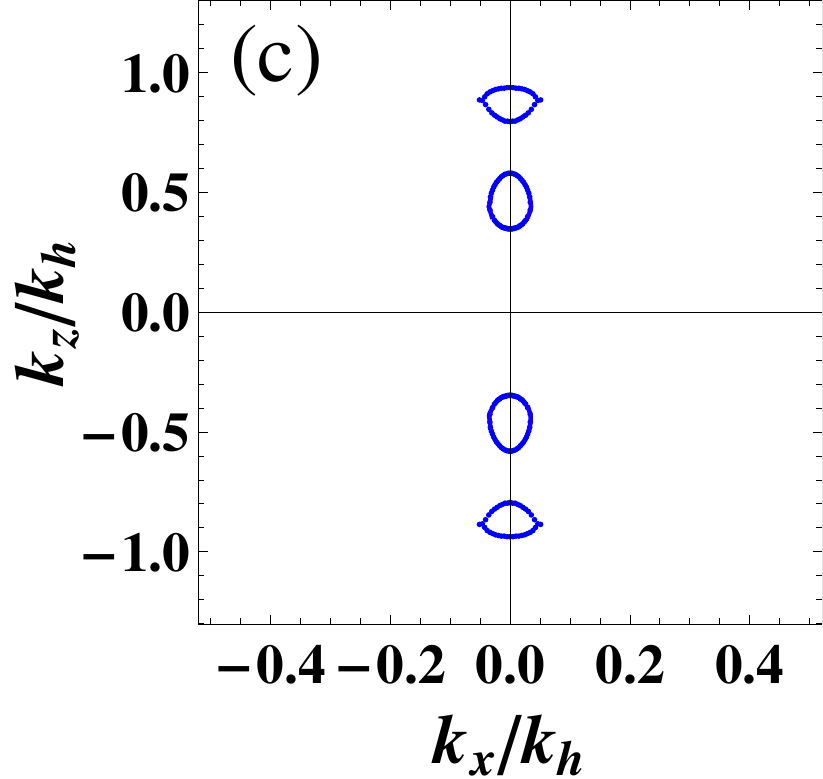}
\includegraphics[width=4.2cm]{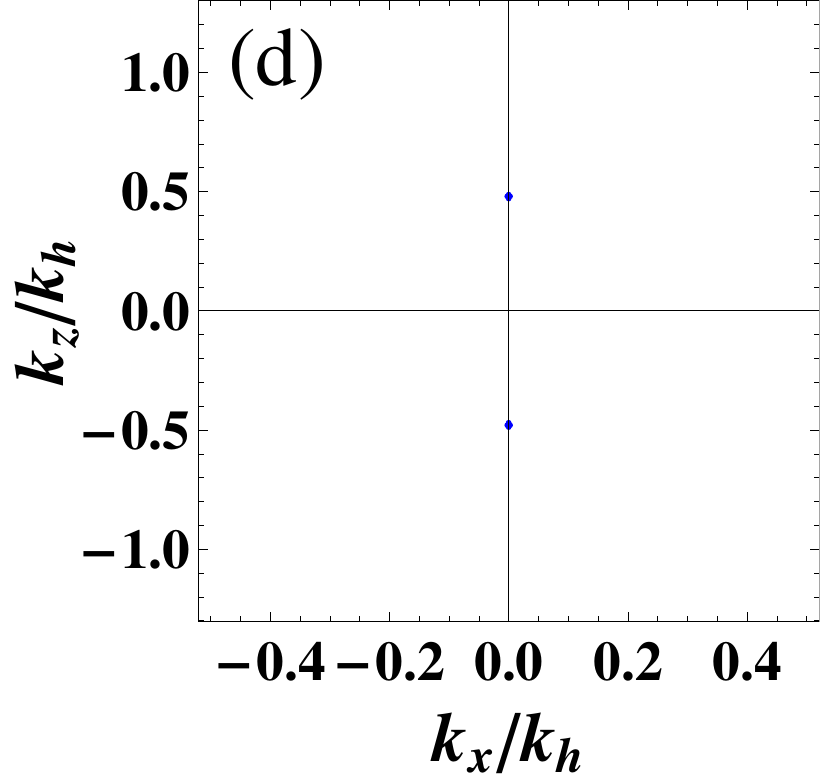}
\includegraphics[width=4.2cm]{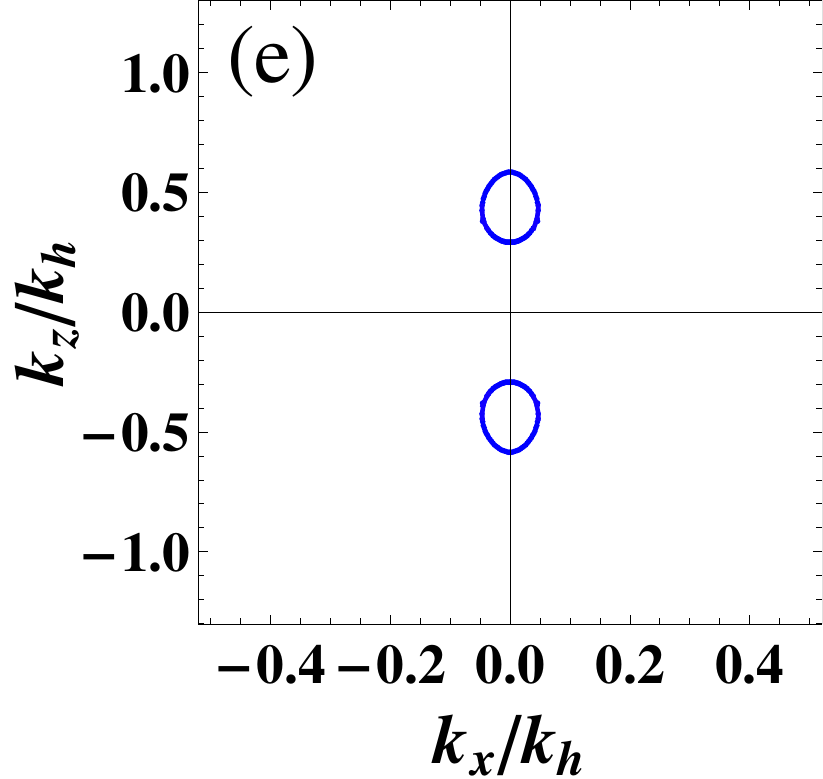}
\includegraphics[width=4.2cm]{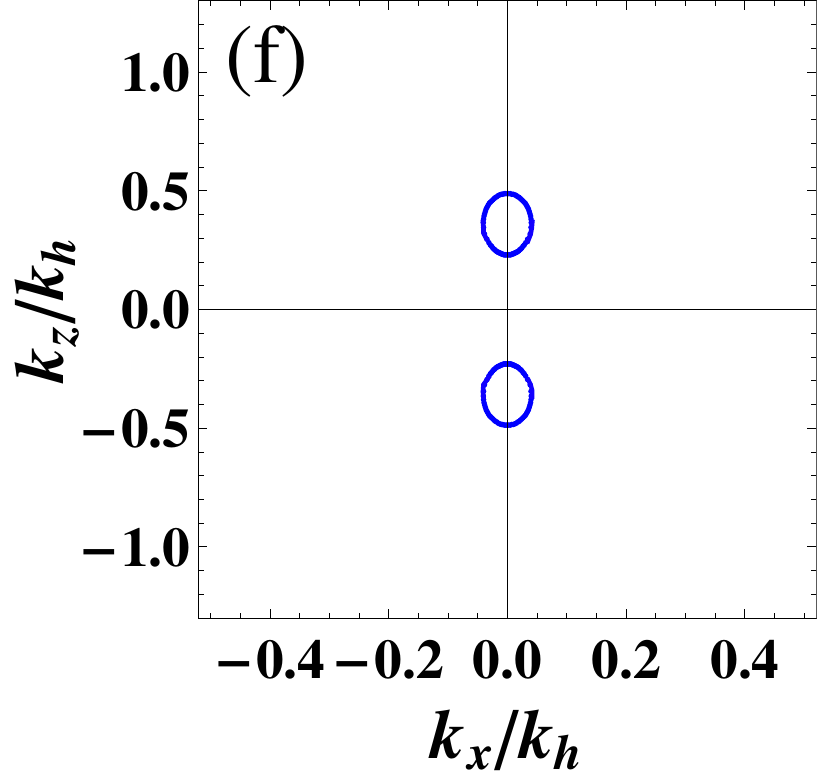}
\caption{(Color online) Typical evolution of gapless contours for nodal FFLO states in the $k_x$--$k_z$ plane ($k_y=0$) with fixed SOC strength $\alpha k_h/h=1.5$ for (a)(c)(e), and $\alpha k_h/h=1.6$ for (b)(d)(f), respectively. The remaining parameters are: (a) $\mu/h=0.16$, $\Delta/h\sim0.36$, $Q_z/k_h\sim-0.21$; (b) $\mu/h=0.05$, $\Delta/h\sim0.41$, $Q_z/k_h\sim-0.20$; (c) $\mu/h=0.2$, $\Delta/h\sim0.38$, $Q_z/k_h\sim-0.21$; (d) $\mu/h=0.09$, $\Delta/h\sim0.43$, $Q_z/k_h\sim-0.20$; (e) $\mu/h=0.22$, $\Delta/h\sim0.39$, $Q_z/k_h\sim-0.20$; (f) $\mu/h=0.15$, $\Delta/h\sim0.46$, $Q_z/k_h\sim-0.19$. For all figures, $\gamma=0.5$, $(k_h a_s)^{-1}=-1$. While (a)(c)(e) represents an evolution between the disconnected nFFLO1 states via an intermediate nFFLO2 state, (b)(d)(f) represents an evolution between the nFFLO1 states across a boundary with only gapless points on the $k_z$ axis. The gapless surfaces are axially symmetric in momentum space with respect to the $k_z$ axis, and the gapless points are symmetric with respect to the $k_z=0$ plane due to the spectra symmetry $E^{+}_{k_z,\nu}=-E^{-}_{-k_z,\nu'}$.}\label{figfflocontour}
\end{figure}

With the understanding of zero center-of-mass momentum pairing states, we now investigate possible FFLO pairing states under 3D SOC and an axial Zeeman field. In the weak coupling limit, it is easy to see that the combination of 3D SOC and the axial Zeeman field makes the Fermi surfaces asymmetric in momentum space along the $\hat{z}$ direction. Furthermore, it has been pointed out before that for systems with SOC and asymmetric Fermi surfaces, the conventional zero center-of-mass momentum pairing state would become unstable against finite center-of-mass momentum FFLO pairing states \cite{chuanweifflo,wyfflo,hufflo}. We will show that this is also the case here. To see this point more clearly, we first perform a small center-of-mass momentum $\mathbf{Q}=(0,0,Q_z)$ expansion of the thermodynamic potential Eq. (\ref{thermoeqn}) around the local minimum with $Q=0$ (see Appendix \ref{sec:appendix-A} for details)
\begin{equation}
\Omega(\Delta,Q_z)=\Omega_0(\Delta)+\Omega_1(\Delta)Q_z+\Omega_2(\Delta)Q_z^2+{\cal O}(Q_z^3).
\end{equation}
We find that for finite pairing order parameter $\Delta$ and axial Zeeman field $h$, the first-order expansion coefficient $\Omega_1$ is typically non-zero and has the opposite sign to $h$. Therefore the local minimum in the thermodynamic potential with $Q=0$ is shifted onto the finite $\mathbf{Q}$ plane. For $h>0$, this suggests that pairing states with zero center-of-mass momentum in the presence of 3D SOC are always unstable against an FFLO pairing state with $Q_z<0$.

\begin{figure}[tbp]
\centering
\includegraphics[width=8cm]{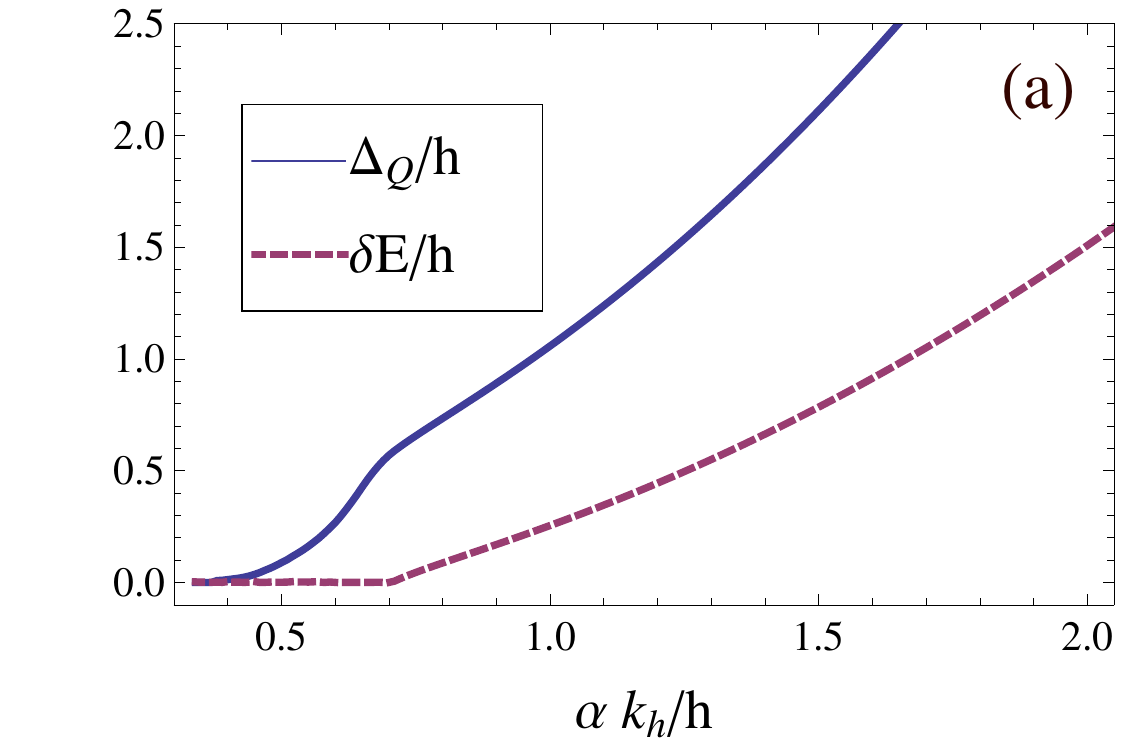}
\includegraphics[width=8cm]{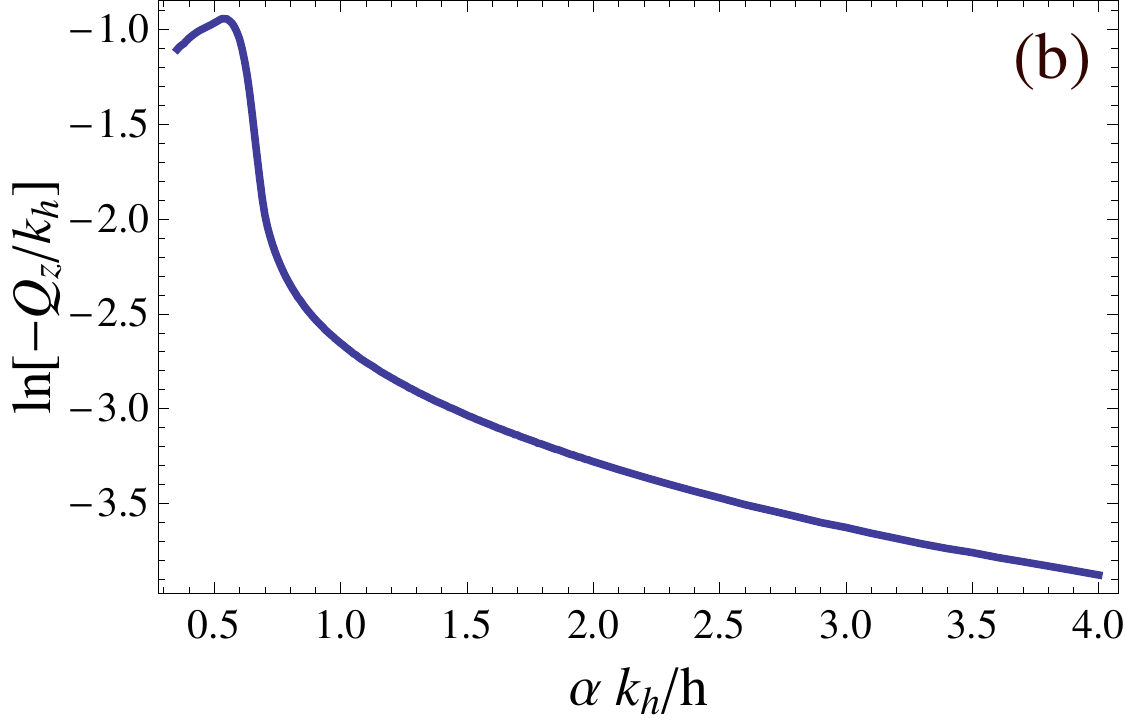}
\caption{(Color online) (a) Evolution of the pairing order parameter $\Delta_{\mathbf{Q}}$ and the minimum excitation gap $\delta E$ of the ground state as a function of SOC strength $\alpha$ for fixed chemical potential $\mu$. (b) Evolution of the logarithm of the center-of-mass momentum of the ground state as a function of SOC strength $\alpha$ for fixed chemical potential. For both figures, we have set $\gamma=1$, $(k_h a_s)^{-1}=0$, $\mu/h=0.2$. }\label{figqzscale}
\end{figure}

\begin{figure}[tbp]
\includegraphics[width=7.6cm]{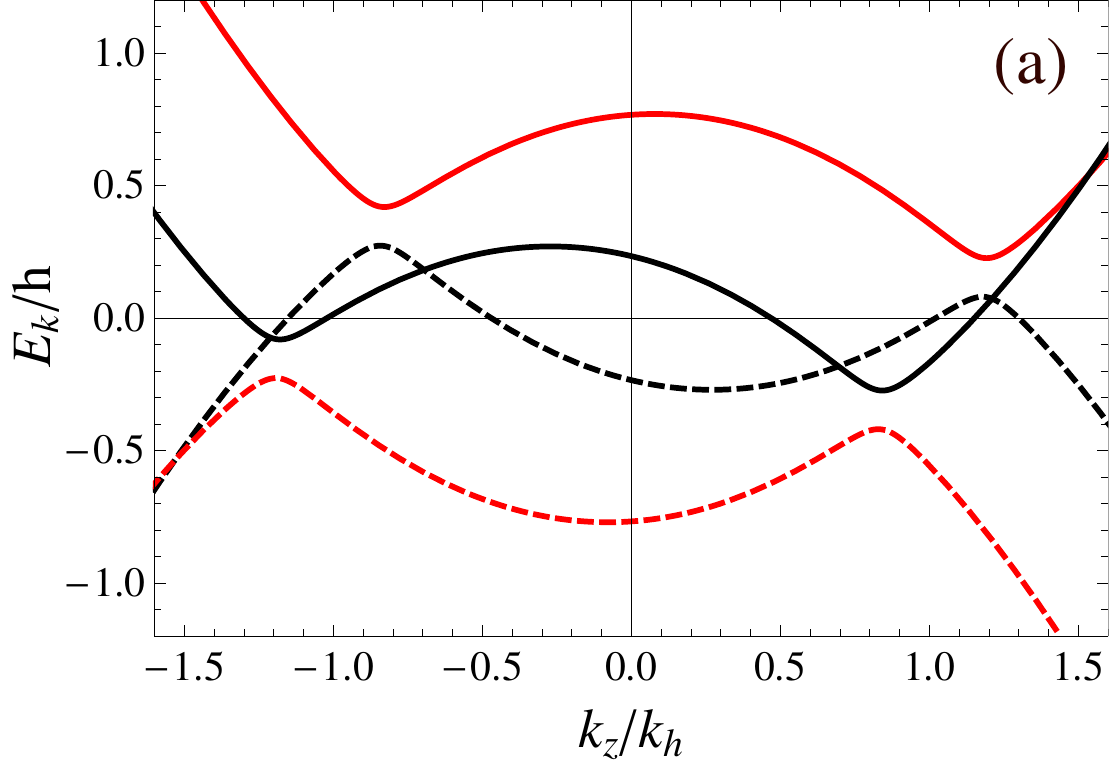}
\includegraphics[width=7.6cm]{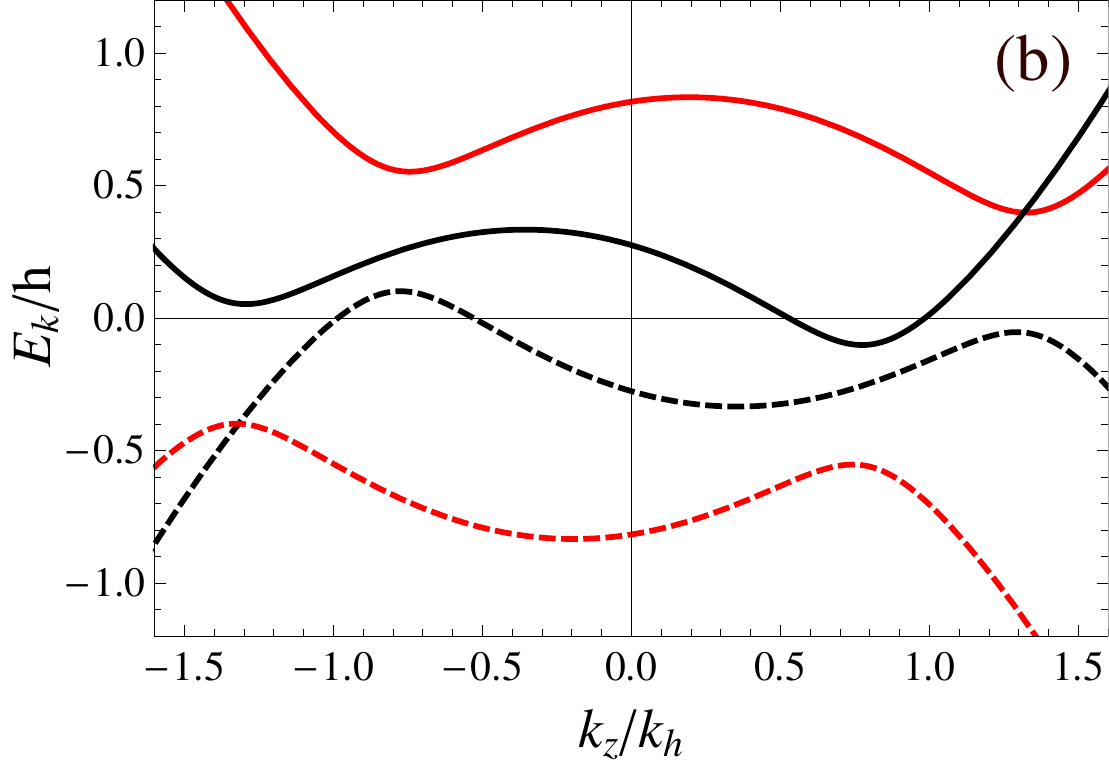}
\includegraphics[width=7.6cm]{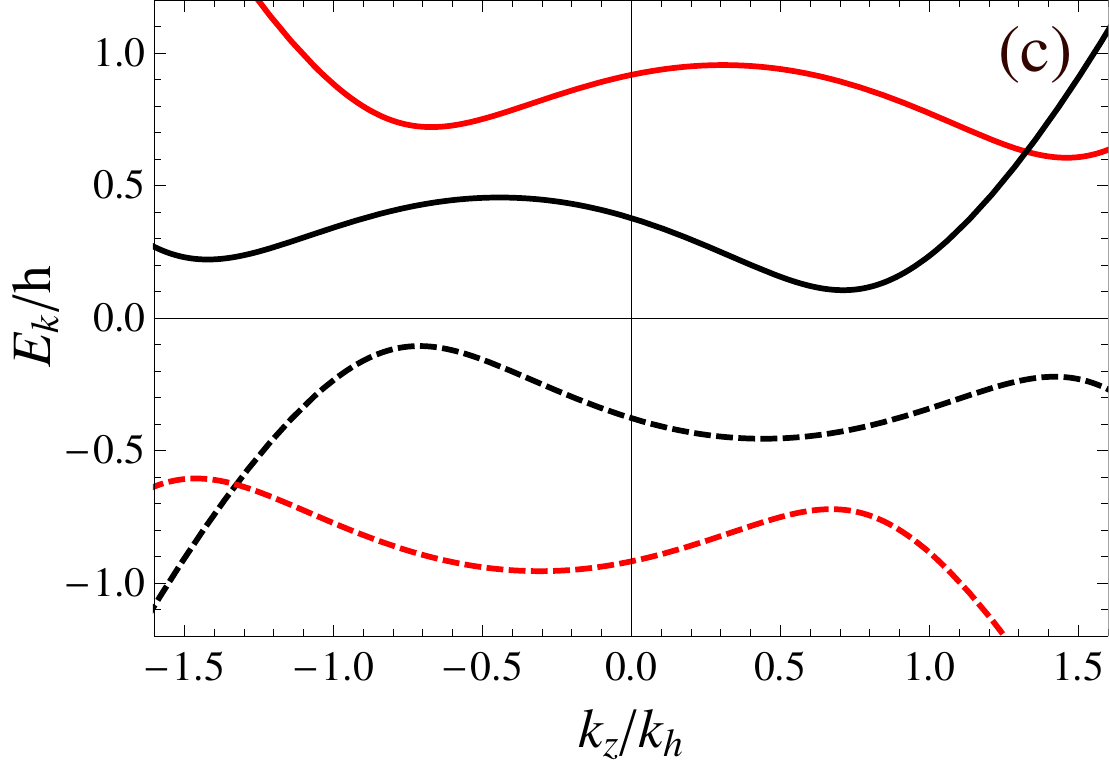}
\caption{(Color online) Typical quasi-particle (solid) and quasi-hole (dashed) dispersion spectra. (a) $\alpha k_h/h=0.7$ (nFFLO2); (b) $\alpha k_h/h=1.1$ (nFFLO1); (c) $\alpha k_h/h=1.5$ (g-FFLO). For all figures, $\mu/h=1$, $\gamma=0.5$, $(k_h a_s)^{-1}=-1$. }\label{figdispersion}
\end{figure}

We then map out typical phase diagrams on the $\mu$--$\alpha$ plane for pairing states with finite center-of-mass momentum (see Fig. \ref{figfflophase}). On the phase diagrams, all the zero center-of-mass momentum pairing states are replaced by FFLO states with center-of-mass momentum opposite to the direction of the axial Zeeman field ($Q_z<0$), as we have expected from previous analysis. Note that for appropriate parameters, pairing states with center-of-mass momentum perpendicular to the Zeeman field are also stable against pairing states with $Q=0$, but are metastable against the ground state on the phase diagram. As we have discussed before, a qualitative understanding here is that the local minima of the thermodynamic potential in the $Q=0$ case are shifted onto the finite $\mathbf{Q}$ plane. When this shift in phase space is not too large, many properties of the resulting FFLO states are similar to those of the original zero center-of-mass momentum pairing states. For instance, we still have first-order phase boundaries between different FFLO states; we still have nodal FFLO states with two (nFFLO1) or four (nFFLO2) gapless surfaces in momentum space (see Fig. \ref{figfflocontour}); and we have a continuous boundary between the nodal FFLO state (nFFLO1) and a fully gapped FFLO state (g-FFLO). Furthermore, there appear to be two disconnected nFFLO1 states on the phase diagram, which are separated by a stability region of the nFFLO2 state for smaller $\alpha$. Across the boundary between nFFLO1 and nFFLO2, typically, a new pair of gapless points appear on the $k_z$ axis and develop into gapless surfaces, which is similar to the evolution of gapless surfaces across the nSF1 and nSF2 boundary in the $Q=0$ case. On the other hand, for larger SOC strength, the nFFLO1 states are typically separated by a continuous boundary, along which the gapless surfaces reduce to two gapless points along the $k_z$ axis. We illustrate the projections of the gapless surfaces in the $k_x$--$k_z$ ($k_y=0$) plane for the two different scenarios in Fig. \ref{figfflocontour}. These FFLO states, the g-FFLO state in particular, are characteristic of pairing states in the presence of SOC and Fermi surface deformation. For example, in the weak coupling limit, one can clearly see that the g-FFLO state features an SOC-induced single-band pairing, which necessarily has finite center-of-mass momentum due to the Fermi surface asymmetry. As the Fermi surface asymmetry decreases with increasing chemical potential $\mu$ and/or SOC strength $\alpha$, we find that the center-of-mass momentum of the g-FFLO state decreases in magnitude (see Fig. {\ref{figqzscale}}(b)). We expect that similar FFLO pairing states should exist in systems with general forms of SOC and Fermi surface asymmetry.

As the difference in gapless surfaces for different FFLO states originate from the quasi-particle(-hole) dispersion spectra (see Fig. \ref{figdispersion}), a straightforward way to identify these different phases experimentally is to measure the quasi-particle dispersion. This can be achieved for example, via the momentum resolved radio-frequency spectroscopy. Alternatively, as different topology of gapless surfaces in momentum space should give rise to different low energy excitations, the rich phases and phase transitions in the system can be probed by measuring the thermodynamic properties.

\begin{figure}[tbp]
\includegraphics[width=8cm]{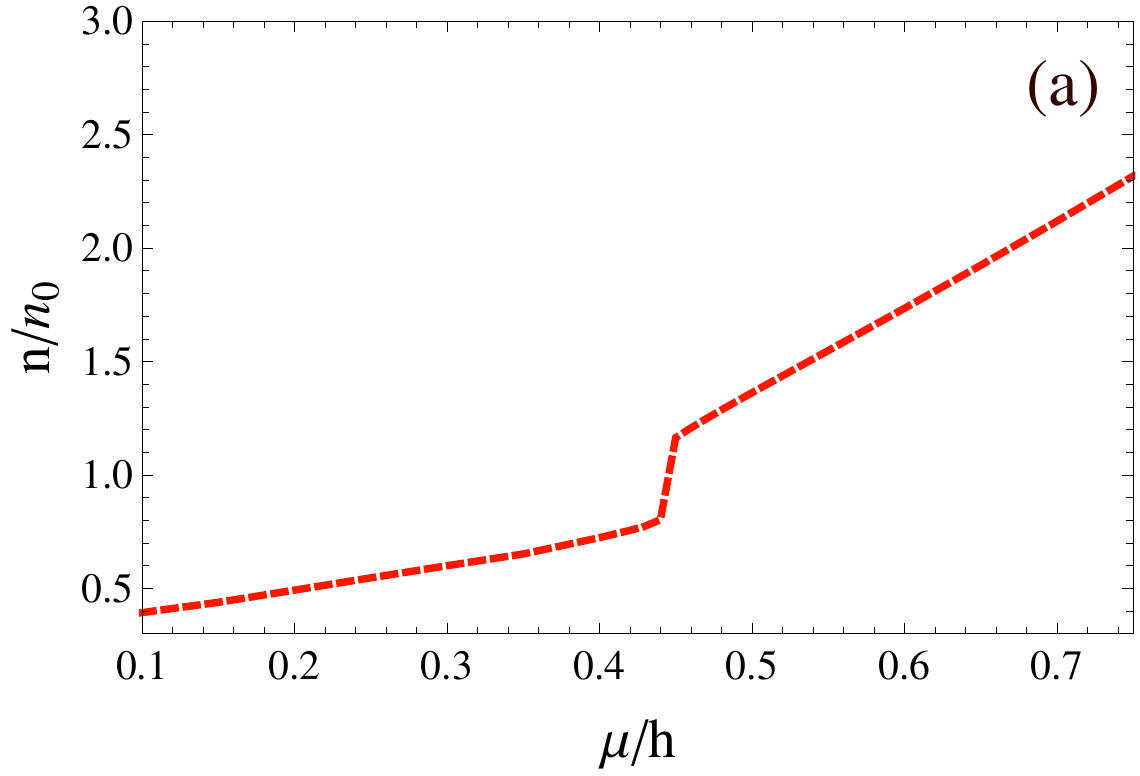}
\includegraphics[width=8cm]{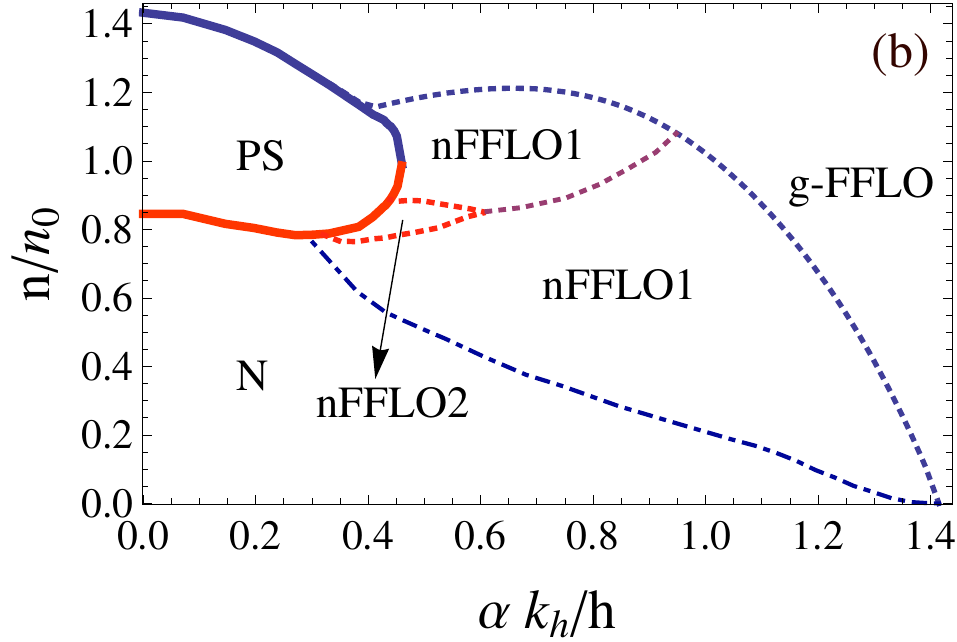}
\caption{(Color online) (a) Total particle number density across the first-order phase boundary for fixed SOC strength $\alpha k_h/h=0.4$. We connect the data points (dots) with dashed lines to guide the eyes. (b) Phase boundary on the $n$--$\alpha$ plane, where the solid curves are first-order boundaries, the dashed curves are the continuous boundaries between different FFLO states, and the dash-dotted curve is the normal state threshold with $\Delta/h=10^{-3}$. For both figures, $\gamma=1$, $(k_h a_s)^{-1}=0$. The unit for the particle density $n_0=k_h^3/(3\pi^2)$.}\label{figNalpphase}
\end{figure}

Finally, we emphasize that due to the existence of first-order phase boundaries on the phase diagram, one must explicitly take phase-separated states into account to get the correct ground state of a uniform gas. As we have discussed previously, in a uniform gas, the total particle density is typically fixed by the number equation. However, across the first-order phase boundary, the total particle density of the ground state does not change continuously (see Fig. \ref{figNalpphase}(a)). This suggests that for certain total number densities, the number equation and the gap equation could not be solved simultaneously, unless a phase separated state is considered. More explicitly, we show in Fig. \ref{figNalpphase}(b) the phase diagram on the $n$--$\alpha$ plane, with the total number density $n=-(1/V)\partial \Omega/\partial\mu$. Here, a phase-separated (PS) region can be clearly identified, where the ground state of the system is a phase-separated state either between the normal state and the g-FFLO state, or between two different FFLO states. This further implies that for calculations in the canonical ensemble, where the total particle number is fixed, one must consider the possibility of phase separation.

\section{Summary}

We have studied the pairing states in a 3D Fermi gas under 3D SOC and an effective axial Zeeman field. We show that the interplay of SOC, Zeeman field and pairing superfluidity lead to rich phase structure with exotic pairing states. In particular, over large parameter regions, the ground state of the system is an FFLO state with finite center-of-mass momentum. These FFLO states exhibit interesting excitation properties, ranging from having different number of gapless surfaces in momentum space to being fully gapped. We argue that these FFLO states are unique to the coexistence of SOC and Fermi surface asymmetry, and are qualitatively different from the conventional FFLO states in a polarized Fermi gas. With the recent proposals for realizing 3D SOC in ultracold atom gases, our work is helpful for the future experimental investigations with 3D SOC in ultracold Fermi gases, and provides valuable information for the general understanding of pairing physics in spin-orbit coupled fermionic systems.


\acknowledgments
We would like to thank Lin Dong, Hui Hu, and Han Pu for helpful discussions. This work is supported by NFRP (2011CB921200, 2011CBA00200), NNSF (60921091), NKBRP (2013CB922000), NSFC (11004186, 11105134, 11274009), SRFDP (20103402120031), the Fundamental Research Funds for the Central Universities (WK2470000001, WK2470000006), and the Research Funds of Renmin University of China (10XNL016). W.Z. would also like to thank the NCET program for support.

\appendix

\section{Instability of BCS state against FFLO pairing}
\label{sec:appendix-A}

In this appendix, we present some details regarding the expansion of thermodynamic potential Eq. (\ref{thermoeqn})
around the BCS state with respect to small center-of-mass momentum $\mathbf{Q}=(0,0,Q_z)$. For the BCS state, we denote the
eigenvalues of the matrix $M_{\bf k}^{Q=0}$ in the effective Hamiltonian Eq. (3) as $E_{{\bf k},\nu}^{\lambda}(Q=0)$,
where $\nu, \lambda = \pm$. By considering a small center-of-mass momentum $\mathbf{Q}=Q_z\hat{z}$, the four quasi-particle (hole)
dispersions can be written as
\begin{eqnarray}
\label{eqn:AppendixA-eigenvalues}
E_{{\bf k},\nu}^{\lambda}(Q_z) = E_{{\bf k},\nu}^{\lambda}(0) + \delta_{1, {\bf k},\nu}^{\lambda} Q_z +
\delta_{2, {\bf k},\nu}^{\lambda} Q_z^2 + {\cal O}(Q_z^3).
\end{eqnarray}
To determine the expansion coefficients $\delta$'s, we employ the identity
\begin{eqnarray}
{\rm tr} \left( A^n \right) = \sum_i \lambda_i^n
\end{eqnarray}
for $n \le 4$, where $\lambda_i$ is the $i$th eigenvalues of matrix $A$.
By matching coefficients of terms involving $Q_z$ and $Q_z^2$ on both sides the equation above,
we can write down two sets of linear equations for $\delta_{1, {\bf k},\nu}^{\lambda}$ and
$\delta_{2, {\bf k},\nu}^{\lambda}$, respectively. After some straightforward algebra,
the expansion of thermodynamic potential Eq. (\ref{thermoeqn}) with respect to small $Q_z$
can be obtained via substitution of the resulting dispersions.


\begin{thebibliography}{99}
\bibitem{footnote1} We note that the implementation of the effective Zeeman field in such a system usually depends on how the synthetic SOC is realized. For the 3D SOC scheme proposed in Ref. \cite{3dsoc1} for instance, the effective Zeeman field can be realized and tuned by adjusting the Rabi-frequencies of the lasers generating the synthetic SOC.
\end{thebibliography}
\end{document}